\documentclass[aps,prd,nofootinbib,preprint,superscriptaddress]{revtex4}  
\usepackage{amsmath,amssymb,exscale}
\usepackage{psfrag,graphicx}
\usepackage[sub,ovp]{psfragx}
\usepackage{overpic}
\usepackage{epsfig}
\usepackage{color,amsmath,amscd,shadow,fancybox,axodraw}
\usepackage[latin1]{inputenc}

\newcommand{\Bt}{\tilde B}
\newcommand{\Wt}{\tilde W}
\newcommand{\qt}{\tilde q}
\newcommand{\ut}{\tilde u}
\newcommand{\dt}{\tilde d}

\newcommand{\Bslash}{\not \!\!B}
\newcommand{\Wslash}{\not \!\!W}
\newcommand{\Btslash}{\not \!\!\Bt}
\newcommand{\Wtslash}{\not \!\!\Wt}
\newcommand{\dslash}{\not \!\partial}
\newcommand{\pslash}{\not \!p}
\def\I{\rm 1\kern-.24em l}

\begin{document}

\begin{flushright}
UdeM-GPP-TH-08-166
\end{flushright}

\title{Electroweak precision constraints on the \\ Lee-Wick Standard Model}

\author{Ezequiel \'Alvarez} 
\affiliation{CONICET and Departamento de F\'{\i}sica, FCEyN, Universidad de Buenos Aires,
Ciudad Universitaria, Pab.1, (1428) Buenos Aires, Argentina} 

\author{Leandro Da Rold}
\affiliation{Instituto de F\'{\i}sica, Universidade de S\~ao Paulo,
Rua do Mat\~ao 187, S\~ao Paulo, SP 05508-900, Brazil} 

\author{Carlos Schat} 
\affiliation{CONICET and Departamento de F\'{\i}sica, FCEyN, Universidad de Buenos Aires,
Ciudad Universitaria, Pab.1, (1428) Buenos Aires, Argentina} 

\author{Alejandro Szynkman} 
\affiliation{Physique des Particules, Universit\'e
de Montr\'eal, C.P. 6128, succ. centre-ville, Montr\'eal, QC,
Canada H3C 3J7} 

\date{\today}

\begin{abstract}
We perform an analysis of the electroweak precision observables in the
Lee-Wick Standard Model. 
%As a result we present constraints on the parameters of the model
%determining the allowed regions in the space of Lee-Wick masses. 
The most stringent restrictions come from the $S$ and $T$ parameters
that receive important tree level and one loop contributions.
In general the model predicts a large positive $S$ and a negative $T$.
To reproduce the electroweak data, if all the Lee-Wick masses are of the
same order, the Lee-Wick scale is of order 5 TeV. 
We show that it is possible to find some regions in
the parameter space with a fermionic state as light as
$2.4-3.5$ TeV, at the price of rising all the other
masses to be larger than $5-8$ TeV. 
%An exception to this result is the mass of the down Lee-Wick fermions, 
%that is not well constrained by the data. 
To obtain a light Higgs with such heavy resonances a
fine-tuning of order a few per cent, at least, is needed. We also
propose a simple extension of the model
including a fourth generation of Standard Model fermions with
their Lee-Wick partners. We show that in this case it is possible to
pass the electroweak constraints with Lee-Wick fermionic masses of
order $0.4-1.5$ TeV and Lee-Wick gauge masses of order 3 TeV. 
\end{abstract}

\maketitle

\section{Introduction}\label{intro}
The Standard Model (SM) describes the electroweak (EW) interactions 
with an incredible precision. However, the instability of the Higgs
potential under radiative corrections signals our ignorance over the
real mechanism of electroweak symmetry breaking (EWSB) and has lead
to many extensions beyond the SM. 
The upcoming LHC era is likely to provide us the tools to check some of the 
proposed solutions to this problem.

Recently, Grinstein, O'Connell and Wise proposed a new extension of
the SM~\cite{Grinstein:2007mp}, 
based  on the ideas of Lee and Wick \cite{Lee:1969fy,Lee:1970iw} for a
finite theory of quantum electrodynamics.
The building block of the Lee-Wick proposal is to consider that the
Pauli-Villars regulator describes a physical degree of freedom. In the
Lee-Wick Standard Model (LWSM), this idea is extended to all the SM
in such a way that the theory is free from quadratic divergences and 
the hierarchy problem is solved. Every SM field has a LW-partner with
an associated LW-mass, these masses are the only new parameters in the
minimal LWSM. A potential problem in this model is that the LW-states
violate causality at the microscopic level due to the opposite sign of their
propagators. However the authors of Ref.~\cite{Grinstein:2007mp} argued
that there is no causality violation on a macroscopic scale provided
that the LW-particles are heavy and decay to SM-particles.
The LWSM can be thought as an effective theory coming from a higher derivative 
theory. %Acausality is suppressed at scales much lower than the LW scale. 
However, to insure perturbative unitarity, higher dimension operators cannot be 
of any type, only those compatible with a LW effective Lagrangian are
acceptable \cite{Grinstein:2007iz}. In ref.~\cite{Grinstein:2007mp} it
was shown with a specific example that unitarity is preserved  due to
the unusual sign of the LW-particles width.
Further considerations on the unitarity of the theory have been presented extensively in the previous literature 
\cite{Cutkosky:1969fq,Nakanishi:1971jj,Lee:1971ix,Nakanishi:1971ky,Antoniadis:1986tu},
the non-perturbative formulation has been discussed in 
\cite{Boulware:1983vw,Jansen:1993jj,Jansen:1993ji,Fodor:2007fn,Knechtli:2007ea} and the one-loop renormalization of LW-gauge theories has been discussed in \cite{Grinstein:2008qq}.

Recent work discussed the suppression of flavor changing neutral
currents \cite{Dulaney:2007dx}, gravitational
LW particles \cite{Wu:2007yd} and the possibility of coupling the
effective theory to heavy particles \cite{Espinosa:2007ny}.
On the phenomenological side, the implications for LHC
\cite{Rizzo:2007ae,Krauss:2007bz} and ILC \cite{Rizzo:2007nf} have
also been discussed. 

The LWSM does not provide any information on the origin of the
LW-masses. However, in order to solve the hierarchy problem these
masses should not be heavier than a few TeV. On the other hand,
the LW particles can not be too light without getting in conflict with
EW precision observables~\cite{lepparadox}. Therefore the aim of our
work is to carry out an
analysis of the electroweak precision tests (EWPT) and derive 
bounds on the masses of the LWSM. 
Since the main motivation to
introduce the LWSM was to solve the hierarchy problem, large LW-masses
will be a source of fine-tuning and will partially spoil the original
motivation. In this way the success of the LWSM is associated to its
efficiency to pass the EWPT without introducing a severe fine-tuning
in the theory.

On the experimental side, determining the parameter space allowed by
the EWPT is crucial to know whether the LWSM could be tested or not
in the next experiments, in particular at the LHC. 

With these motivations we have performed an analysis of the EW
observables in the LWSM. As in the original
formulation~\cite{Grinstein:2007mp}, we have assumed the principle of
minimal flavor violation (MFV) to simplify the flavor structure of the
model. The most stringent constraints come from the
$S$ and $T$ Peskin-Takeuchi parameters~\cite{oblique}. We present our
results as lower bounds on different combinations of the LW-masses
of the gauge and quark fields%: $M_1,M_2,M_q,M_u,M_d$
. If we assume
degenerate LW-masses for all the fields, the LW-scale allowed by
the EWPT
%, as given by the $95\%$ Confidence Level 
%fit of the Electroweak Working Group, 
is of order 5 TeV and there is little chance to test this model at the
LHC. Relaxing the constraint on equal masses, it
is possible to find configurations in the parameter space where one
of the masses can be as low as $2.4-3.5$ TeV, at the price of rising the
other masses to be $\gtrsim 5-8$ TeV. This situation is more favorable
from the experimental side and it might be accessible at the LHC.

Concerning the fine-tuning of the model, a heavy Higgs gives further
contributions to $S$ and $T$ pointing in the same direction as the
contributions from the LW-fields, and for this reason is
strongly disfavoured. Thus, in order to obtain a light Higgs one has to 
cancel the rather large contributions to the Higgs mass from the
LW-particles running in the loops, that are proportional to the
LW-masses squared. We estimate the needed tuning of
the model to be at least of order a few per cent. 

A possible way to relax the constraints from the EWPT
would be to generate an extra positive contribution to $T$
without increasing at the same time the $S$ parameter.
By including a fourth generation of fermions of SM-type, with their 
corresponding LW-partners, it is possible to generate a large $T$, 
without generating a too large $S$. 
To obtain a $T$ parameter of the needed size one has
to assume an approximate custodial symmetry for the Yukawas of 
the fourth generation. 
We show that for vector LW-masses of order 3 TeV and Yukawa
masses of the fourth generation in the range $0.2-1.2$ TeV, it is
possible to have all the fermionic LW-masses in the range $0.4-1.5$
TeV and pass the EWPT.

The paper is organized as follows. In Sec.~\ref{model} we give a very
brief description of the LWSM, in Secs.~\ref{tree} and~\ref{loop} we
compute the tree and radiative contributions to the EW precision
parameters. In Sec.~\ref{analysis} we scan over the parameter space of
the model and present a detailed analysis of the allowed regions. 
We consider the extension of the LWSM by including a fourth 
generation in section \ref{4gen}.
We conclude in Sec.~\ref{conclusions} and show the details of some of the
calculations in the Appendices.

\section{The LWSM}\label{model}
The LWSM was originally formulated introducing a higher derivative term
for each of the SM-fields. The theory contains one new parameter for
every SM-field, the LW-mass corresponding to the dimensional
coefficient of the associated higher derivative term. The authors of
Ref.~\cite{Grinstein:2007mp} introduced new LW-fields and showed that
it is possible to reformulate the theory in terms of these fields in
such a way that there are no higher derivative terms in the
Lagrangian. In this formulation the LW-masses are the masses of the
LW-fields and, although the LW-fields mix with the SM ones, 
the particle content of the theory is more
transparent. The LW-fields have the same quantum numbers as their SM
partners and the couplings between the SM and LW-fields are the same
as the SM couplings, although the signs of the interactions are not
always the usual ones. 
It is possible to consider even higher derivative terms ({\it{e.g.}}
six-derivative terms) that will in general lead to additional
LW-states, however we will not consider this case.
We refer the reader to
Ref.~\cite{Grinstein:2007mp} for the details and quote here some
specific interaction terms that are useful to understand the
contributions to the EW observables. We will denote the fields
associated to the LW-states with a tilde.

The quadratic Lagrangian for the gauge SM and LW-fields, after setting
the Higgs to its vacuum expectation value (VEV), is: 
\begin{eqnarray}\label{eqGauge2}
\mathcal{L}_{2g}=&-&\frac{1}{2}{\rm{tr}}\left(B_{\mu\nu}B^{\mu\nu}-\Bt_{\mu\nu}\Bt^{\mu\nu}
+W_{\mu\nu}W^{\mu\nu}-\Wt_{\mu\nu}\Wt^{\mu\nu}\right)
\nonumber\\
&-&\frac{1}{2}(M_1^2\Bt_\mu\Bt^\mu+M_2^2\Wt^a_\mu\Wt_a^\mu)
+\frac{g_2^2v^2}{8}(W^{1,2}_\mu+\Wt^{1,2}_\mu)(W_{1,2}^\mu+\Wt_{1,2}^\mu)\nonumber\\
&+&
\frac{v^2}{8}(g_1B_\mu+g_1\Bt_\mu-g_2W^{3}_\mu-g_2\Wt^{3}_\mu)(g_1B^\mu+g_1\Bt^\mu-g_2W_3^\mu-g_2\Wt_3^\mu),
\end{eqnarray}
where $W_{\mu\nu}=\partial_\mu W_\nu-\partial_\nu W_\mu, W_\mu=W_\mu^a T^a$ and similar
for the other fields, and $g_{1,2}$ are the hypercharge and weak
couplings. The sign of the kinetic and mass terms of the LW-fields are
opposite to the usual ones. This sign is responsible for the cancellation of
the quadratic divergences as well as the microscopic causality
violations by the LW-particles.

The quadratic Lagrangian for the fermionic fields after setting the
Higgs to its VEV is:
\begin{eqnarray}\label{eqFermion2}
\mathcal{L}_{2\psi}&=&\sum_{\psi=q_L,u_R,d_R}\bar\psi i\dslash\psi
-\sum_{\psi=q,u,d}\bar{\tilde\psi}(i\dslash-M_\psi)\tilde\psi\nonumber\\
& &-m_u(\bar{u}_R-\bar{\tilde{u}}_R)(q_L^u-\tilde q_L^u)
-m_d(\bar{d}_R-\bar{\tilde{d}}_R)(q_L^d-\tilde q_L^d)+\rm{h.c.},
\end{eqnarray}
where a generation index is understood, $q^t=(q^u,q^d)$ denotes the
SU(2)$_L$ doublet, $m_{u,d}=\lambda_{u,d} v/\sqrt{2}$ and for simplicity we
have omitted the leptonic sector. 
Different to the SM chiral fermions, the LW-fermions combine into
Dirac spinors of masses $M_{q,u,d}$. 
We will assume that the LW-fermions transforming in the
same representation of the gauge symmetries have the same mass, this
is compatible with the MFV principle~\cite{MFV}. 
Then the matrices $M_\psi$ of
Eq.~(\ref{eqFermion2}) are proportional to the identity and we will
trade $M_\psi\rightarrow{\I} M_\psi$, with $M_\psi$ a scalar
parameter. For effects on FCNC when MFV
is not satisfied see Ref.~\cite{Dulaney:2007dx}.

The quadratic Lagrangian for the Higgs field is:
\begin{equation}
\mathcal{L}_{2H}=(\partial_\mu H)^\dagger(\partial^\mu
H)-(\partial_\mu \tilde H)^\dagger(\partial^\mu \tilde
H)+M_h^2\tilde H^\dagger \tilde H-\frac{m_h^2}{2}(h-\tilde
h)^2\; ,
\end{equation}
where $H^t=(h^+,\frac{v+h+iP}{\sqrt{2}})$ and 
$\tilde H^t=(\tilde h^+,\frac{\tilde h+i\tilde P}{\sqrt{2}})$,
$m_h^2=\lambda v^2/2$ and $M_h$ is the LW-mass. Only the physical
Higgs field $h$ and its LW-partner $\tilde h$ mix.

The gauge fermionic interactions are: 
\begin{eqnarray}\label{eqLint}
\mathcal{L}_{int}=&-&\sum_{\psi=q_L,u_R,d_R}\![g_1\bar\psi(\Bslash+\Btslash)\psi
+g_2\bar\psi(\Wslash+\Wtslash)\psi]\nonumber \\
&+&\sum_{\psi=q,u,d}\left[g_1\bar{\tilde\psi}(\Bslash+\Btslash)\tilde\psi
+g_2\bar{\tilde\psi}(\Wslash+\Wtslash)\tilde\psi\right].
\end{eqnarray}
Note that the LW-fermions couple to the gauge fields with the opposite
sign compared with the SM-fermions.

For LW-mass scales much larger than the top mass the mixings between the light
SM-fermions and the LW-fermions can be neglected. For this reason only
the third generation will contribute to the EW precision
parameters. In Appendix~\ref{appendixMf} we diagonalize the fermionic
mass matrix. In Eqs.~(\ref{fmasses}-\ref{frotation}) we show the
physical masses and the matrices connecting the flavor and mass basis
for $M_q\neq M_{u,d}$. The case for $M_q= M_{u,d}$ has been considered
in Ref.~\cite{Krauss:2007bz}.

Finally we want to comment on the naturalness of the model. The
authors of Ref.~\cite{Grinstein:2007mp} showed explicitly that the gauge
one loop contributions to the Higgs mass are only logarithmically
sensitive to the cut-off of the theory. We want to stress that this
contribution is proportional to the square of the vector LW mass, $M_g$, $\delta
m_h^2\simeq \frac{3C_2(N)g^2}{16\pi^2}M_{g}^2
\log\frac{\Lambda^2}{M_{g}^2}$, in such a way that the quadratic
divergence is recovered when the LW-mass is divergent. The same effect
is present in the fermionic contribution to the Higgs mass, 
$\delta m_h^2\simeq \frac{N_c \lambda^2}{8\pi^2}M_{f}^2
\log\frac{\Lambda^2}{M_{f}^2}$, with $M_f$ the fermionic
LW-mass. Therefore, to have a light Higgs in
a natural way, the LW-vectors (fermions) should be lighter than $\sim
2$ TeV ($\sim 600$ GeV), with a mild dependence on the cut-off $\Lambda$.

\section{Tree level contributions to the EW precision parameters}\label{tree}
We discuss in this section the tree level contributions to the EW
precision parameters. We will show that the only parameters that
are important in the LWSM are the oblique parameters $S$ and $T$. In
the next section we will compute the 1-loop corrections to $S$ and $T$
and show that the radiative contributions can be as large as the tree
level ones.

In the LWSM the mixings between the gauge bosons and their
LW partners induce non-canonical couplings for
the SM fermions %(see Eq.~(\ref{eqGauge2}))
\footnote{See Ref.~\cite{Agashe:2003zs} for a
discussion of this effect in another context.}.
A shift in the gauge fermion couplings can
be reabsorbed into the oblique parameters. Therefore, to correctly define
the oblique parameters $S,T,U$ it is necessary to properly normalize
the couplings between the fermions and the gauge bosons~\footnote{This
observation was overlooked in Ref.~\cite{Grinstein:2007mp}.}. We find it
useful to work in the effective theory obtained after integrating out
the heavy LW fields at tree level. Setting the Higgs field to its VEV,
the interactions in the effective theory that are important to
normalize the gauge fermion couplings are:
\begin{eqnarray}\label{acoupling}
\mathcal{L}_{eff}=&-&g_2 W^{\mu\;1}J_\mu^1\left[1-\frac{g_2^2 v^2}{g_2^2v^2-4M_2^2}\right]-g_2 W^{\mu\;2}J_\mu^2\left[1-\frac{g_2^2 v^2}{g_2^2v^2-4M_2^2}\right]
\nonumber \\
&-&J_\mu^3\left[g_2 W^{\mu\;3}-(g_2 W^{\mu\;3}-g_1 B^\mu)\frac{g_2^2 v^2 M_1^2}{g_1^2 v^2
M_2^2+(g_2^2v^2-4M_2^2)M_1^2}\right]\nonumber \\
&-& J_\mu^Y\left[g_1 B^\mu-(g_1 B^\mu-g_2 W^{\mu\;3})\frac{g_1^2 v^2 M_2^2}{g_1^2 v^2
M_2^2+(g_2^2v^2-4M_2^2)M_1^2}\right]
\end{eqnarray}
where $J_\mu^i$ are the usual currents of SM fermions, and we have
considered that the momentum of the LW-vectors is small, $p^2\ll M_i^2$. Since the coefficients in
Eq.~(\ref{acoupling}) are the same for all the
generations, the same redefinition of the gauge fields leads to
canonical gauge couplings for all the SM fermions:
\begin{eqnarray}
\mathcal{L}_{eff}=-g_2 \sum_{a=1,2,3}W^{\mu\;a}J_\mu^a-g_1
B^{\mu}J_\mu^Y \; .
\end{eqnarray}
The gauge kinetic and mass terms induce contributions to the oblique parameters
after the gauge field redefinitions. The tree level contributions to
$S$ and $T$ are:
\begin{eqnarray}
S &=& 4 \pi v^2 \left( \frac{1}{M_1^2} + \frac{1}{M_2^2}
\right)+\mathcal{O}\left(\frac{v^4}{M_i^4}\right)\; ,
\label{Stree}\\
T &=& \pi \frac{g_1^2 + g_2^2}{g_2^2} \frac{v^2}{M_1^2}\; . \label{Ttree}
\end{eqnarray}
Eq.~(\ref{Ttree}) is valid to all order in $v$ in the tree level approximation.
Moreover, notice that the sign difference between 
Eq.~(\ref{Ttree}) and the result of Ref.~\cite{Grinstein:2007mp} is due to the additional contribution coming from the redefinition of the gauge fields mentioned above.
We can see that for $M_1\rightarrow \infty$ the tree level $T$ parameter
cancels, as expected since in this limit we partially recover a custodial
symmetry in the LW-gauge sector.

The $U$ parameter is of order
$\mathcal{O}\left(\frac{v^4}{M_i^4}\right)$ and will be neglected in
our analysis.

The effective Lagrangian also includes four fermion operators
generated by exchange of LW vector fields, with coefficients of order 
$g_i^2/(2M_i^2)$. The constraints from these operators are
weaker than the constraints from the oblique parameters.

The mixings of SM and LW fermions also induce non-canonical couplings,
this effect could be important for the $b$-quark.
The mixings between $b_L$ and the LW fermions are of order $m_b/M_q$. 
Therefore to protect the $Zb_L\bar b_L$ coupling that is in agreement
with the SM prediction at the $0.25\%$, it is enough to have a LW mass
$M_q\geq \mathcal{O}(100\rm{GeV})$. On the other hand, 
the experimental measurements of
the forward-backward asymmetry of the bottom quark indicate a
deviation in the $Zb_R\bar b_R$ coupling of order $25\%$, $\delta
g_R^b\sim 0.02$ . Since in the LWSM the $b_R$
mixings are of order $m_b/M_d$, to generate the required anomalous
coupling at tree level we would need a very low mass $M_d\sim
\mathcal O$(10 GeV), already excluded. 
\footnote{To agree with the experimental data a $\delta
g_R^b\sim 0.17$ is also possible, but it would require even
lighter new particles.}

\section{Radiative contributions to $S$ and $T$}\label{loop}
In this section we compute the one-loop contributions to the oblique
parameters $S$ and $T$. The most important contributions come from the
third generation of LW-fermions.

\subsection{Contributions to $T$ from the gauge-Higgs sector}\label{Thiggs}
\begin{figure}
\begin{center}
\begin{picture}(300,60)(0,0)
	\Vertex(-45,25){3}
	\Vertex(5,25){3}
	\Text(-20,60)[c]{$\tilde H$}
	\Text(-20,-10)[c]{$\tilde H$}
	\Photon(-70,25)(-45,25){3}{3}
	\Photon(5,25)(30,25){3}{3}
	\DashArrowArc(-20,25)(25,0,180){4}
	\DashArrowArc(-20,25)(25,180,360){4}
	\Text(-65,-10)[c]{(a)}
	\Vertex(90,25){3}
	\Vertex(140,25){3}
	\Text(115,60)[c]{$H$}
	\Text(115,-10)[c]{$\tilde A$}
	\Photon(65,25)(90,25){3}{3}
	\Photon(140,25)(165,25){3}{3}
	\DashCArc(115,25)(25,0,180){4}
	\PhotonArc(115,25)(25,180,360){3}{6}
	\DashLine(90,25)(80,50){4}
	\DashLine(140,25)(150,50){4}
	\Text(70,-10)[c]{(b)}
	\Vertex(240,25){3}
	\Text(240,60)[c]{$\tilde H$}
	\Photon(215,25)(240,25){3}{3}
	\Photon(240,25)(265,25){3}{3}
	\DashCArc(240,38)(12,0,360){4}
%	\DashCurve{(240,25)(230,40)(240,50)(250,40)(240,27)}{4}
	\Text(240,-10)[c]{(c)}
\end{picture}
\caption{One-loop Feynman diagrams contributing to the oblique
parameters involving the Higgs sector.}
\label{figLWHiggs}
\end{center}
\end{figure}
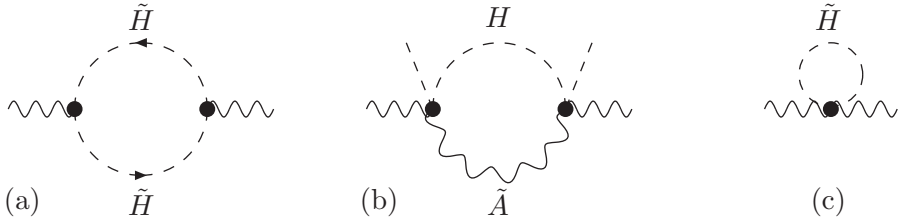

The one-loop Feynman diagrams involving the LW-Higgs field $\tilde H$
are shown in Fig.~\ref{figLWHiggs}. We discuss first the contributions to $T$.
There is no custodial symmetry in the LWSM protecting the $T$
parameter. Thus there is no reason to expect finite radiative
contributions to $T$. As expected from the general arguments of 
Ref~\cite{Grinstein:2007mp}
there are no quadratic divergences, however, we obtain corrections
from the LW-Higgs sector that are logarithmically sensitive to the 
UV cut-off of the theory.
%that involve the Feynman diagrams with zero external momentum. 

We consider the different diagrams of Fig.~\ref{figLWHiggs} in some
detail. Since the charged and pseudoscalar LW-Higgs fields do not mix with their SM
partners, the diagrams (a) and (c) of Fig.~\ref{figLWHiggs} cancel in
the combination $\Pi_{11}-\Pi_{33}$ and do not contribute to
$T$. 

The diagrams of Fig.~\ref{figLWHiggs}(b) with one SM-Higgs and one
LW-gauge field can be explicitly computed. For $M_{1,2}\gg m_W^2$ we
can perform an expansion in Higgs-VEV insertions. The leading
contribution comes from the zeroth order term, {\it{i.e.}}: we neglect the
mixings of the LW-gauge fields due to the mass insertions. In this
limit the Feynman diagrams give 
$\Pi_{11}(0)-\Pi_{33}(0)\simeq
\frac{g_1^2g_2^2v^2}{64\pi^2}\frac{m_h^2}{M_1^2}\log\frac{\Lambda^2}{m_h^2}$.
A brief explanation of this result is as follows: 
there is a factor $g_1g_2v/2$ for each vertex, the factor $1/(16\pi^2)$
comes from the loop and the sign is different from the SM-Higgs
contribution because the LW propagator has an extra minus sign. Again,
this contribution to $T$ cancels for infinite $M_1$.

From the previous result we obtain $T\simeq \frac{g_1^2+g_2^2}{4\pi}\frac{m_h^2}{M_1^2}
\log\frac{\Lambda^2}{m_h^2}$, that is logarithmically divergent with the
cut-off. However, for a light Higgs and LW-gauge
masses larger than $\sim2$~TeV, these contributions are smaller than the tree
level ones, Eq.~(\ref{Ttree}), even in the limit of $\Lambda\sim
M_{Pl}$. As we will show in the next section, they
are also smaller than the fermionic contributions.

There are contributions to $T$ from the diagram of
Fig.~\ref{figLWHiggs}(c), replacing the LW-Higgs propagator by a
LW-gauge one. At leading order in a mass insertion expansion, this
contribution exactly cancels because the SU(2)$_L$ LW-gauge fields are
degenerate. There is a non-vanishing contribution at second order but it
is suppressed by a factor $m_W^2/M_2^2$, and can be neglected.

\subsection{Contributions to $S$ from the gauge-Higgs sector}\label{Shiggs}
We discuss the LW-Higgs sector contributions to $S$ from the
Fig.~\ref{figLWHiggs}. 
%The Feynman diagram of Fig.~\ref{figLWHiggs}(c)
%does not contribute to $S$ because there is no external momentum
%circulating in the loop.
For $m_h\ll M_h$ all the LW-Higgs components are degenerate, thus at
leading order in a mass insertion expansion the Feynman diagrams
corresponding to Fig.~\ref{figLWHiggs}(a) cancel out. The first
non-trivial contribution is due to the splitting between the neutral
LW-Higgs and the other LW-Higgs components. This gives a small
$S\simeq\frac{m_h^2}{24\pi M_h^2}$.  

The Feynman diagram of Fig.~\ref{figLWHiggs}(b) gives a small
contribution also, 
$S\simeq\frac{1}{2\pi}\left(\frac{m_W^2}{M_2^2}+\frac{m_Z^2 s^2}{M_1^2}\right)$, with $s=\sin \theta_W$.

This contributions to $S$ can be neglected compared with the tree
level one, Eq.~(\ref{Stree}).

\subsection{Fermionic contributions to $T$}\label{fermionicT}
\begin{figure}
\begin{center}
\begin{picture}(300,80)(0,0)

	\Vertex(-45,45){3}
	\Vertex(5,45){3}
	\Text(-20,80)[c]{$\psi$}
	\Text(-20,10)[c]{$\psi'$}
	\Photon(-70,45)(-45,45){3}{3}
	\Photon(5,45)(30,45){3}{3}
	\ArrowArc(-20,45)(25,0,180)
	\ArrowArc(-20,45)(25,180,360)
	\Text(-65,0)[c]{(a)}

	\Vertex(270,45){3}
	\Vertex(320,45){3}
	\Vertex(276,61){3}
	\Vertex(313,61){3}
	\Vertex(276,29){3}
	\Vertex(313,29){3}
	\Text(295,80)[c]{$u,\tilde u$}
	\Text(295,10)[c]{$u,\tilde u$}
	\Text(282,45)[c]{$q,\tilde q$}
	\Text(308,45)[c]{$q,\tilde q$}
	\Photon(245,45)(270,45){3}{3}
	\Photon(320,45)(345,45){3}{3}
	\ArrowArc(295,45)(25,0,180)
	\ArrowArc(295,45)(25,180,360)
	\DashLine(278,60)(259,80){4}
	\DashLine(312,60)(331,80){4}
	\DashLine(276,30)(259,10){4}
	\DashLine(312,30)(331,10){4}
	\Text(250,0)[c]{(c)}

	\Vertex(120,45){3}
	\Vertex(170,45){3}
	\Vertex(145,70){3}
	\Vertex(145,20){3}
	\Photon(95,45)(120,45){3}{3}
	\Photon(170,45)(195,45){3}{3}
	\ArrowArc(145,45)(25,0,90)
	\ArrowArc(145,45)(25,90,180)
	\ArrowArc(145,45)(25,180,270)
	\ArrowArc(145,45)(25,270,360)
	\DashLine(145,70)(145,90){4}
	\DashLine(145,20)(145,0){4}
	\Text(130,75)[c]{$q,\tilde q$}
	\Text(130,15)[c]{$q,\tilde q$}
	\Text(165,75)[c]{$u,\tilde u$}
	\Text(165,15)[c]{$u,\tilde u$}
	\Text(100,0)[c]{(b)}

\end{picture}
\caption{One-loop Feynman diagrams contributing to the oblique
parameters involving the fermionic sector. Diagram (a) is the
fermionic loop in the mass basis. Diagrams (b,c) are the first non-trivial
contributions from the up sector expanding in Yukawa mass insertions.}
\label{figLWfermions}
\end{center}
\end{figure}
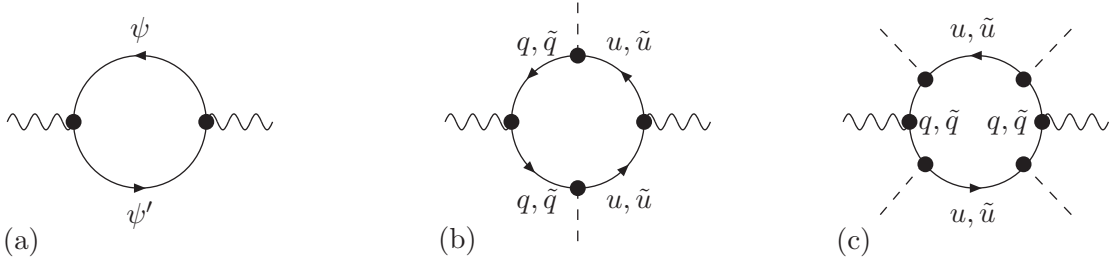

The $T$ parameter measures the amount of isospin breaking, thus the
third generation, having the largest Yukawa couplings, gives the
dominant contribution compared with the other fermions.
We will show that the fermionic contributions of
Fig.~\ref{figLWfermions} dominate also over the
other loop corrections. 
To check our calculations we have computed them in two different ways. We refer
the reader to the appendices for the details. 

One way to compute the fermionic $T$ is by working in the diagonal
mass basis. Inserting the rotation matrices $S_{L,R}^{u,d}$, defined in
Eqs.~(\ref{frotationl}) and~(\ref{frotation}), into the gauge fermion interactions,
Eq.~(\ref{eqLint}), we
obtain the couplings between the fermions and the SM-gauge fields in
the mass basis.
Since there are no mixings in this basis, we just
have to sum over all the possible fermionic combinations in the loop
diagram of Fig.~\ref{figLWfermions}(a).
The matrices $S_{L,R}^{u,d}$
have been calculated in a perturbative mass insertion expansion, then
the results obtained by this method are valid for $m_{u,d}\ll
M_{u,d,q}$. To obtain a non vanishing $T$ one has to consider at
least four mass insertions, this implies that we have to expand
$S_{L,R}^{u,d}$ to that order.
The result is almost independent on $M_d$ because the
down Yukawa is much smaller than the up Yukawa for the third
generation. 

For small LW-fermionic masses ($M_{q,u}\gtrsim m_t$) the convergence
of the mass insertion series is rather slow. In fact, for masses
$M_{q,u}\lesssim 1$ TeV we have checked that the first non-trivial
contribution in the perturbative expansion has large deviations from
the non-perturbative one, and can not be trusted. For this reason we
will consider also the resummation of the mass insertion series. The diagram of Fig.~\ref{figLWfermions}(c) gives
the first non-trivial contribution to $T$ in the flavor basis in the
mass insertion expansion. For $\Pi_{33}$, the
fermionic propagators (at zeroth order in mass insertions) attached to the gauge vertex
could be either $q_L$ or $\tilde q$, and the fermionic propagators
between the Yukawas could be $u_R$ or $\tilde u$ ($d_R$ or $\tilde d$)
for the up (down) contribution. There is a similar diagram for
$\Pi_{11}$. Using the results of Appendices~\ref{Approp} and \ref{Appvacuum} it is possible to
obtain the fermionic vacuum polarizations to all orders in the mass
insertions. The result for $\Pi_{11}(0)-\Pi_{33}(0)$ is:
\begin{eqnarray}\label{eqTfermion}
& &\Pi_{11}(0)-\Pi_{33}(0)=\frac{g_2^2N_c}{4}\int \frac{d^4p}{(4\pi)^2}
\left[ \frac{p^6-4p^4M_q^2+p^2M_q^2}{(p^2-M_q^2)^4}+\frac{1}{p^2}-
2\frac{p^2-2M_q^2}{(p^2-M_q^2)^2}\right]\times \\
& &\left[
\frac{m_u^2M_u^2(p^2-M_q^2)}{p^4(M_u^2-p^2)+M_q^2(m_u^2M_u^2+p^2(p^2-M_u^2))}-
\frac{m_d^2M_d^2(p^2-M_q^2)}{p^4(M_d^2-p^2)+M_q^2(m_d^2M_d^2+p^2(p^2-M_d^2))}\right]^2
\;
\nonumber
\end{eqnarray}
where $m_{u,d}$ stand for the masses of the third
generation. 
Eq.~(\ref{eqTfermion}) includes the contribution from the
SM-fermions, that must be subtracted to obtain $T$. 
This term is obtained by taking the limit of infinite
LW-masses.

The resulting $T$ parameter is negative and it increases for small LW-masses.
We make an analysis of the results and its
consequences for the LHC in section~\ref{analysis}.

\subsection{Fermionic contributions to $S$}\label{fermionicS}
Perturbatively the fermionic $S$ parameter counts the number of active fermions in
the EW sector. However, at one loop,  doublets $(N,E)$ of chiral
fermions contribute with $S\sim 1/(6\pi)[1-2Y\log(m_N^2/m_E^2)]$,
whereas for vector-like fermions the constant term is absent,
$S\sim 2Y/(3\pi)\log(m_N^2/m_E^2)$. The LW-fermions are
vector-like and the isospin splitting within a doublet is due to the
mixings with the SM fields through the Yukawa couplings. Therefore, for
$M_{q,u}\gg m_t$ we expect the LW-fermions to induce a small $S$ at
loop level. However, due to the mixings between the SM and
LW-fermions, the contributions to $S$ are not so simple, and for
$M_{q,u}\sim m_t$ the
 isospin splitting could be large. 

The fermionic one loop Feynman diagrams contributing to $S$ are shown in
Fig.~\ref{figLWfermions}.
We have computed them using the two methods of
section~\ref{fermionicT}, by working in the diagonal mass basis and
also resumming the mass insertions in the flavor basis. 
For $M_{q,u}\gtrsim 1.5$ TeV the exact one-loop calculation computed
in the flavor basis and the
perturbative calculation in the mass basis agree very well. However,
for $M_{q,u}\lesssim 1$ TeV the perturbative result has large
deviations from the full result. We have checked that including higher
order terms in the mass insertion expansion the convergence is
improved for low values of $M_{q,u}$.
We will use the vacuum polarization resumming the mass
insertions in our analysis (an expression similar to
Eq.~(\ref{eqTfermion}), but much longer, 
can be obtained also in this case, however we
omit it for the sake of brevity). For some details on this
calculations see the Appendix~\ref{Appvacuum}.
The important result is that the fermionic $S$ is negative and
small compared with the tree level $S$ of Eq.~(\ref{Stree}).

\section{Analysis of the EWPT}\label{analysis}
We make a numerical analysis of the LWSM by scanning over the
parameter space of the model: ${M_q,M_u,M_1,M_2}$.
%, where $M_{q,u}$ are|the LW-fermionic masses of the third generation. 
As we argue in section~\ref{fermionicT} the dependence on $M_d$ is negligible (we
have also checked by a numerical calculation that this is true), thus from now
on we fix $M_d=1$ TeV. As argued in sections~\ref{Thiggs} 
and~\ref{Shiggs}, our
results are almost independent on $m_h$ and $M_h$ provided that the Higgs
is light, $m_h\sim 114$~GeV, and the LW-Higgs is sensibly heavier than the
SM-Higgs. In any case, as we will show, a heavy SM-Higgs is
strongly disfavored by the EWPT.

\begin{figure}[h]

$\begin{array}{ccc}
\begin{overpic}[width=7.5cm]{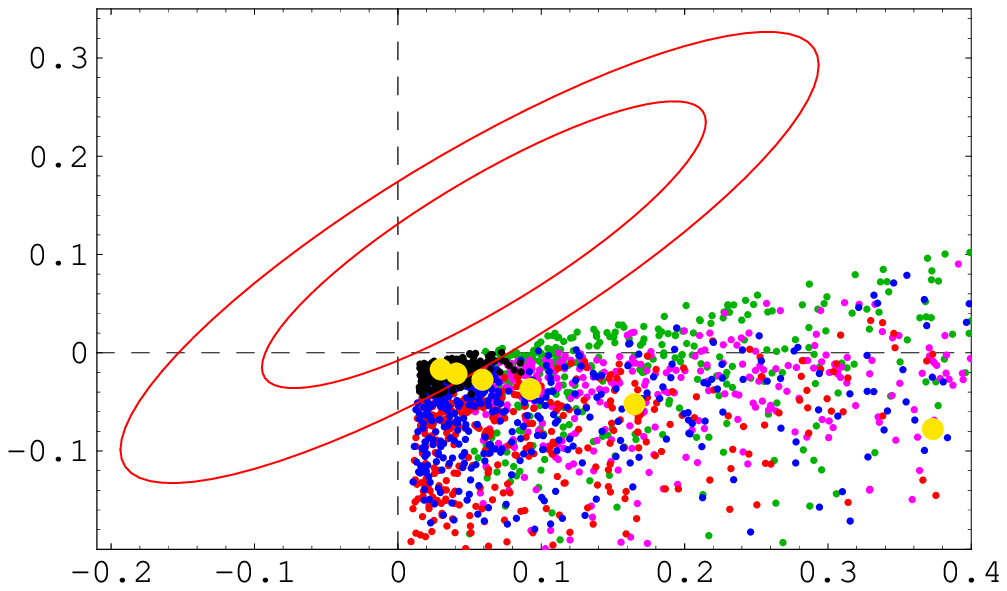}

\put(-7,30){\rotatebox{90}{$T$}}
\put(50,-5){$S$}

\end{overpic} & \ \ \ \ \ &
\begin{overpic}[width=7.5cm]{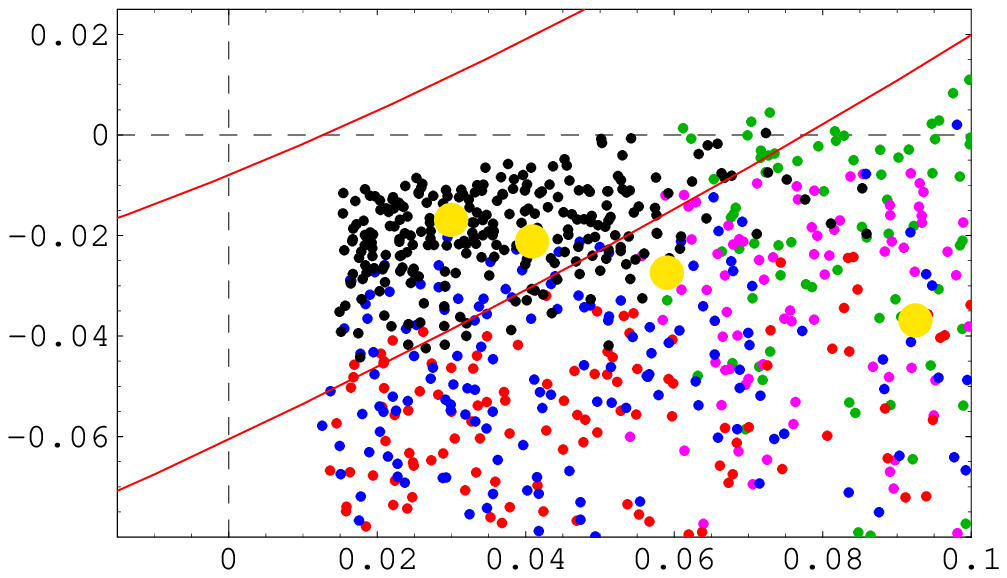}

\put(-7,30){\rotatebox{90}{$T$}}
\put(50,-5){$S$}

\end{overpic}
\end{array}$

\vspace{0.2cm}
\caption{\protect \small $68\%$ and $95\%$ Confidence Level contours
in the $(S,T)$ plane, and LWSM predictions. The black dots indicate
points where all four masses $M_1$ , $M_2$,  $M_q$  , $M_u$ are larger
than 4 TeV. Coloured points correspond to cases where at least one mass
is less than 4 TeV. The colour indicates which mass 
is below 4 TeV: green, magenta, red, blue dots correspond to 
$M_1$ , $M_2$,  $M_q$  , $M_u$ less than 4 TeV, respectively. 
The yellow dots correspond to taking all masses equal
and 7,6,5,4 ... TeV, from left to right. 
\label{planeST}}
\end{figure}

To obtain a better understanding of the importance of the $S$ and
$T$ parameters in constraining the model we show in Fig.~\ref{planeST}
the $68\%$ and $95\%$ Confidence Level contours in the $(S,T)$ plane,
as obtained from the LEP Electroweak Working
Group~\cite{EWWG}, together with
the LWSM predictions for several values of the LW-masses. It is
clear from Fig.~\ref{planeST} that there is no region in the parameter
space lying within the $68\%$ Confidence Level contour (we have considered LW-masses not larger than 10 TeV). There is
however a small but sizable region of the $95\%$ Confidence Level
contour that is covered. Choosing all the LW-masses to be equal
corresponds to the large yellow points in Fig.~\ref{planeST}. 
In this case only masses above 5 TeV enter into the allowed
region. The other coloured points in Fig.~\ref{planeST} correspond to
one of the LW-masses being light (lighter than 4 TeV), whereas the
black dots are for all the LW masses being heavier than 4 TeV. 
We can see that most of the configurations with light
new particles do not satisfy the EWPT, whereas most of the
configurations that pass the EWPT do not have any light new particle.
\vspace{0.1cm}

It is also evident from Fig.~\ref{planeST} that most of the
configurations have a too large positive $S$ and negative $T$
parameters. The positive $S$ contribution is mainly generated at tree level
by the non-canonical fermionic gauge couplings, see Eq.~(\ref{Stree}). The
$T$ parameter has a tree level positive contribution,
Eq.~(\ref{Ttree}), and a negative one-loop contribution due to the third
generation of LW-fermions. For light LW-fermions the one-loop
fermionic correction dominates over the tree level one resulting in a
negative $T$. A light LW-vector could cancel the large negative $T$
generated by the fermions, but generating at the same time a too large
$S$.

We quote now the minimum values of LW-masses that pass the EWPT. 
The lightest $M_1$ ($M_2$) lying inside the $95\%$ contour is
$M_1\simeq 3.2$ TeV (3.8 TeV), and corresponds to a green (magenta) point in
Fig.~\ref{planeST}. 
For a green (magenta) point to lie inside the ellipse, 
$M_2,M_q,M_u$ ($M_1,M_q,M_u$) have to be heavier than 
$\sim 5.4,5.2,3.6$ TeV ($\sim 7.0,8.1,6.0$ TeV), respectively. 
%The lightest $M_2$ inside the ellipse is $M_2\simeq 3.8$ TeV, with 
%$M_1,M_q,M_u$ heavier than $\sim 7,8,6$ TeV respectively. 
The lightest $M_q$ ($M_u$) inside the ellipse is $M_q\simeq 3.5$ TeV 
($M_u\simeq 2.4$ TeV). 
For a red (blue) point to lie inside the ellipse, 
$M_1,M_2,M_u$ ($M_1,M_2,M_q$) have to be heavier than
$\sim 7.5,7.0,4.9$ TeV ($\sim 3.9,4.7,4.2$ TeV), respectively.

A heavy SM-Higgs gives an extra negative $T$ and a positive
$S$~\cite{oblique}. Thus it points in the wrong direction and gives
stronger constraints for the LWSM.
As shown in sections~\ref{Thiggs} and~\ref{Shiggs}, the contributions
from the LW-Higgs can not alleviate this situation.

\begin{figure}[h]
$\begin{array}{ccc}
\begin{overpic}[width=7.cm]{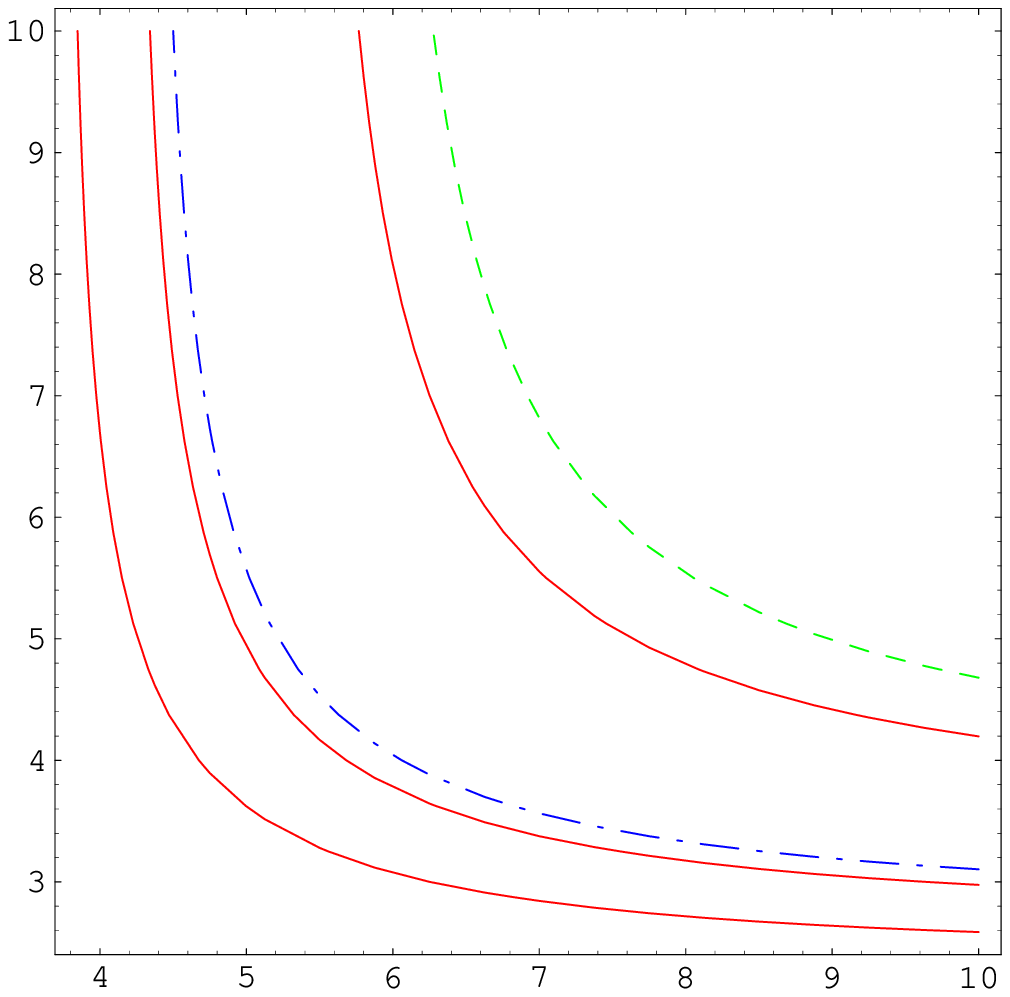}

\put(-5,40){\rotatebox{90}{$M_u$[TeV]}}
\put(40,-5){$M_q$[TeV]}

\end{overpic} & \ \ \ \ \ & 
\begin{overpic}[width=7.cm]{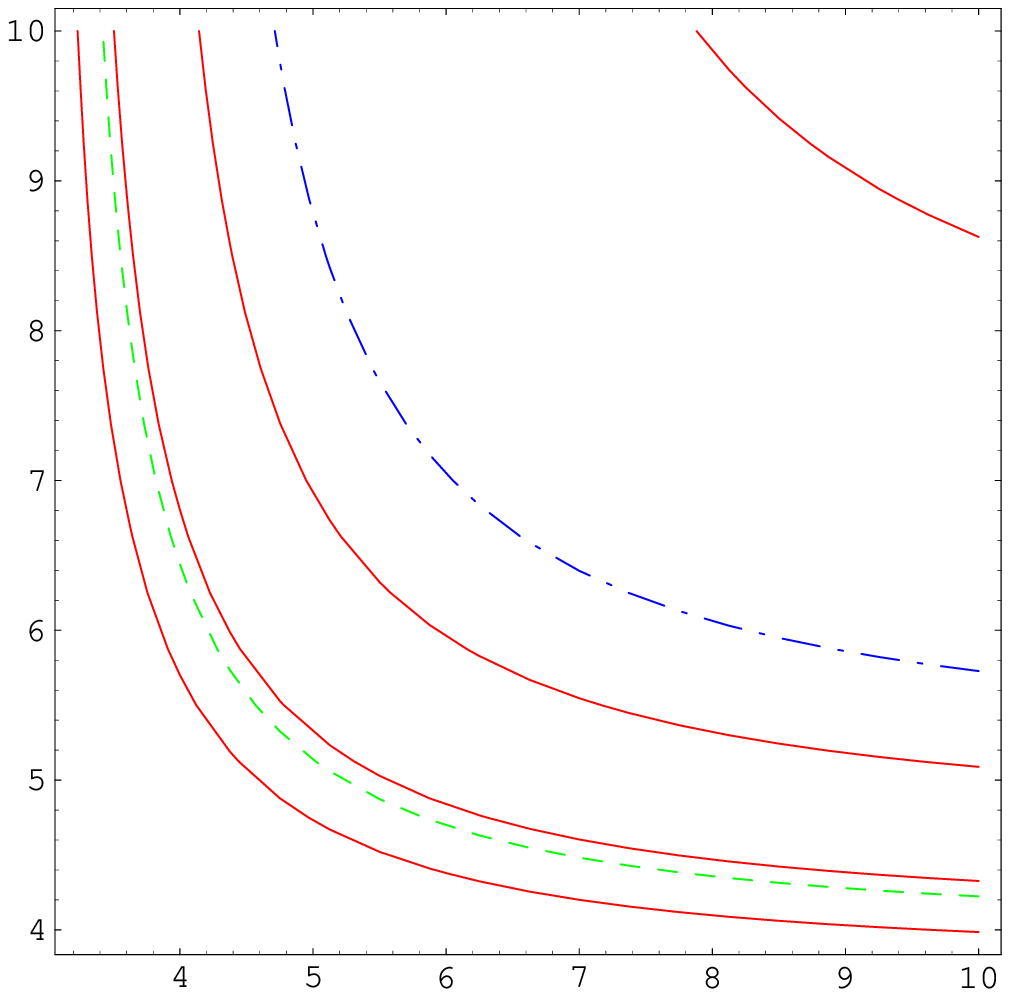}

\put(-5,40){\rotatebox{90}{$M_2$[TeV]}}
\put(40,-5){$M_1$[TeV]}

\end{overpic} 
\end{array}$

\vspace{0.3cm}
\caption{Values of the LW-masses
allowed by the EWPT ($95\%$ Confidence Level contour). The region above the
lines is allowed by the EWPT and the region below the lines is
forbidden. On the left we show the plane $(M_q,M_u)$ for fixed values
of the LW-vector masses. The red lines correspond (from left to right) to $M_1=M_2=7,6,5$ TeV. The
dashed green line to $M_1=10 , M_2=4$ TeV, and the dot-dashed blue
line to $M_1=4 , M_2=10$ TeV. 
On the right we show the plane $(M_1,M_2)$,
for fixed LW-fermionic masses. The red lines correspond (from left to right) to $M_q=M_u=7,6,5,4$ TeV. The
dashed green line to $M_q=10$ TeV and $M_u=4$ TeV, and the dot-dashed blue 
line to $M_q=4$ TeV and $M_u=10$ TeV.
}
\protect \small
\label{planeM}
\end{figure}

In Fig.~\ref{planeM} we show the LW-fermionic masses 
$M_q,M_u$ allowed by the EWPT for fixed values of $M_{1,2}$. We have
considered the $95\%$ confidence level constraints on the $S,T$
parameters. The lines divide the parameter space in an upper region allowed
by the EWPT and a lower region that does not pass the EWPT.
\vspace{0.2cm}

In order to obtain a rather light LW-fermion, for example an SU(2)$_L$
singlet, $\tilde u$, with a mass of order $2.5-3$ TeV, we are forced
by the EWPT to have heavy LW-vectors and also a heavy LW-fermion doublet,
$\tilde q$, with masses larger than $\sim 5$ TeV. 

In Fig.~\ref{planeM} we show also the LW-vector masses 
$M_1,M_2$ allowed by the EWPT for fixed values of $M_{q,u}$.
The regions above (below) the lines (do not) pass the EWPT. To
obtain a light vector the other vector and the fermions are forced to
be heavy, with masses larger than $\sim 5-6$ TeV. In any case the
LW-vector masses can not be lighter than 3 TeV. The lightest vectors are
slightly heavier than the lightest fermions, as they give larger
contributions to $S$. In general the LW-vector $\tilde
B$ can be lighter than $\tilde W$. This is because,
given a positive $S$, the EWPT prefer a positive $T$, that is
generated by $\tilde B$ and not by $\tilde W$.

\section{Extending the LWSM with a fourth generation}\label{4gen}
We consider in this section a very simple extension of the LWSM that
can provide positive contributions to $T$ and a rather small $S$. 
We add a fourth generation of fermions with the same quantum numbers
and chiralities as the ordinary SM generations, together with their
LW-partners. The important terms in the Lagrangian for the 
EWPT are still described by
Eqs.~(\ref{eqFermion2}) and~(\ref{eqLint}), with a generation index
including the fourth generation. For simplicity we will consider that
$M_\psi$, acting on a space of dimension four, is still proportional
to the identity. Therefore, the only new parameters are the Yukawa
couplings of the fourth generation. We will consider only the effects
of the quarks of the fourth generation, moreover, it is very simple to
include the leptons to cancel the anomalies. Ignoring for the moment the
mixings between the SM-fermions and the fourth generation, the new
physical effects are contained in $m_{u,d}^4=\lambda_{u,d}^4\,
v/\sqrt{2}$. 

As explained in section~\ref{fermionicS}, a generation of SM-quarks
with a rather small isospin splitting produces a $S\sim 0.1$, whereas
vector-like quarks do not produce $S$ in the limit of no isospin
splitting. Moreover, for a rather large isospin splitting the $S$
parameter generated by SM-quarks decreases and the $S$ generated by
vector-like quarks remains very small, $S\lesssim 0.04$ for
$m_N\lesssim 2m_E$. Therefore, taking into account the results for the
minimal LWSM, an extra small contribution to $S$ can be
consistent with the EWPT if there is also a small and positive
contribution to $T$.
\begin{figure}[h]
$\begin{array}{c}
\begin{overpic}[width=9cm]{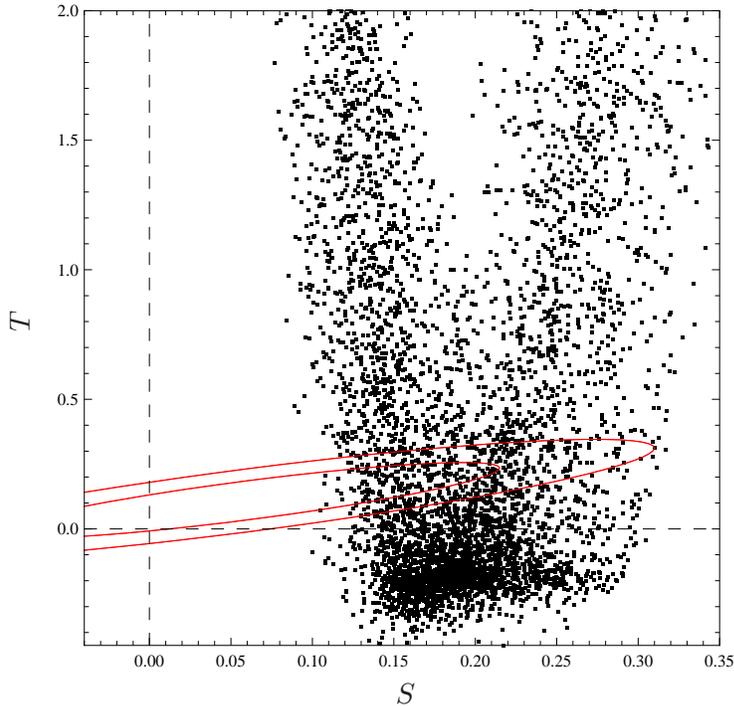}
\put(-7,50){\rotatebox{90}{$T$}}
\put(50,-5){$S$}
\end{overpic}
\end{array}$
\vspace{0.2cm}
\caption{$68\%$ and $95\%$ Confidence Level contours
in the $(S,T)$ plane, and predictions of the LWSM with a fourth
generation. The vector LW-masses are fixed to 3 TeV and $M_q\simeq
1.5$ TeV. The Yukawas and fermionic LW-masses take values in the range
$0.2-1.5$ TeV.}
\label{fig4ta}
\end{figure} 

The $T$ parameter generated by new fermions is proportional to the
isospin splitting of the new sector. Thus the splittings in the Yukawas
of the fourth generation and in the LW-sector produce contributions to
$T$.  Since the mass of the down quark of the fourth generation can be large, $T$
has a strong dependence with $M_d$ in this case.
The $T$ parameter generated by a fourth generation with
their LW-partners, without constraints in the isospin violation, will
be in general of order 1, much larger than the needed $T$.
We have checked that this is indeed the situation in the present
proposal. Therefore, to generate a positive $T$ of the appropriate size, it is
necessary to constrain the isospin splitting.

It is immediate to extend the results of sections~\ref{fermionicT} 
and~\ref{fermionicS} to include the one loop effects of a fourth
generation.
We have scanned over the parameter space fixing the vector LW-masses to
be of order $\sim 3$ TeV, in order to suppress the large tree level
contributions to $S$ and $T$. It is found that a heavy $M_q\sim 3-4$ TeV is 
preferred by the data, but a lower $M_q\sim 1.5$ TeV is still
consistent with the EWPT for light $m_{u,d}^4$ and $M_{u,d}$.

In Fig.~\ref{fig4ta} we show the $68\%$ and $95\%$ Confidence Level
contours in the $(S,T)$ plane together with the predictions of the
LWSM with a fourth generation. We have considered Yukawas and fermionic
LW-masses in the range $0.2-1.5$ TeV. The first thing one can notice is that a
much larger region of the ellipses in the $(S,T)$ plane is covered in this
case, compared with the Fig.~\ref{planeST} of the minimal LWSM. Also
there is a rather large range of values of $T$ covered, as expected
if there is no restriction in the isospin splitting. A heavier $M_q$,
$\sim 3$ TeV, increases $T$, allowing a larger overlap between the
dense region and the ellipses.

\begin{figure}[h]
$\begin{array}{c}
\begin{overpic}[width=7.5cm]{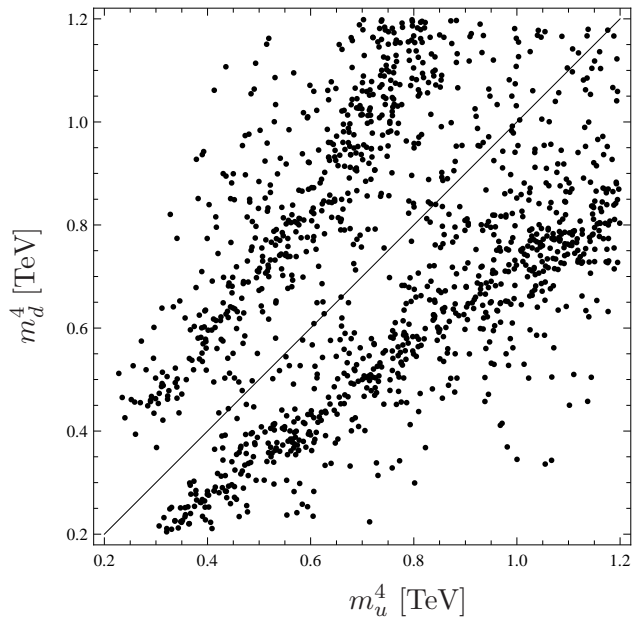}
\put(-10,40){\rotatebox{90}{$m_d^4$ [TeV]}}
\put(50,-7){$m_u^4$ [TeV]}
\end{overpic}
\end{array}$
\vspace{0.3cm}
\caption{Points in the $(m_u^4,m_d^4)$ plane satisfying the EWPT at the $95\%$ Confidence Level in the LWSM with a fourth generation. The line corresponds to $m_u^4=m_d^4$.}
\label{fig4ta2}
\end{figure} 
In Fig.~\ref{fig4ta2} we show the regions in the plane $(m_u^4,m_d^4)$ preferred by the EWPT. We find that the isospin violation due to the Yukawas of the fourth generation has to be rather small, satisfying the approximate relation 
$\frac{|m_u^4-m_d^4|}{m_u^4+m_d^4}\sim 0.3$.
On the other hand, since the effect from the LW-fermions is much smaller, the constraints in the isospin splitting are much weaker for this sector.
We find that there are no regions excluded in the $(M_u,M_d)$ plane, provided that $M_u$ and $M_d$ are larger than $\sim 0.4$ TeV.
\section{Conclusions}\label{conclusions}

We have performed a careful scan over the parameter space of the
minimal LWSM and have determined the regions
that pass the EWPT. The most stringent constraints come from the $S$
and $T$ Peskin-Takeuchi parameters. In particular, the most important
restrictions come from the tree level contributions to $S$ and $T$ and
from  the one-loop fermionic contribution to $T$. 

We have shown that it is necessary to choose very specific values of
the LW-masses to obtain light LW particles that could eventually be
discovered at the LHC. The masses of the vectorial LW-particles are always of
order 3.2 TeV or heavier, with the lightest value obtained at the
price of rising the masses of the fermionic LW-particles to be at
least of order $6-8$ TeV. The fermionic spectrum of LW-particles can be
somewhat lighter than the vectorial one, and it is possible to have
fermions with masses as low as 2.4 TeV. This can be done rising
the LW-masses of the other fermions and vectors to be at least of
order $5-8$ TeV. Whether these heavy states could be produced and
detected at the LHC deserves a careful study, some analysis has been
done in Refs.~\cite{Rizzo:2007ae,Rizzo:2007nf}.

The only possible exception to the previous bounds may be the
down LW-fermion, whose mass is not well constrained. 
Since the bottom Yukawa is small, the EWPT do not give
important restrictions on $M_d$. 
Although the model does not explain the origin of the
LW-masses, we can expect that the same mechanism gives masses to all 
the LW-fermions. In this case $M_d$ may be of the same order as the
other fermionic masses $M_u$ and $M_q$.

The degree of tuning of the model depends on the scale $\Lambda$
where new physics beyond the LWSM shows up. For degenerate LW-masses,
and in the most favorable case, with a small 
$\Lambda\sim\mathcal{O}$(10 TeV), we
estimate a degree of tuning that is at least of order a few per cent
(see the last paragraph of section~\ref{model}).
For larger $\Lambda$ the tuning becomes of order a few per mille. 
In the scenario where a little hierarchy in the LW-spectrum is
allowed, the degree of fine-tuning increases to order a few per mille
already with a small $\Lambda$. Thus, although the LWSM can solve
the hierarchy problem by cancelling the quadratic divergences of the
SM, to pass the EWPT with its minimal version one has to reintroduce
some degree of tuning. 

The constraints from the EWPT can be relaxed extending the minimal LWSM
in such a way that there is an extra positive contribution to $T$
without increasing much, at the same time, the $S$ parameter.
We have shown that this can be done including a fourth generation of
fermions with its LW-partners.
Without any assumption in the isospin violation of the fourth
generation Yukawas and in the LW-fermionic sector, the generated $T$
is too large. Our study shows that the effect of isospin violation
from the Yukawas is larger than the effect from the LW-fermions.
To generate the appropriate $T$ one has to impose an approximate
custodial symmetry for the Yukawas. 
The amount of isospin violation required by the
data is of order $30\%$. We have considered vector LW-masses
of order 3 TeV to suppress the tree level $S$ and $T$, and we have
shown that in this case the fermionic LW-masses can be as small as
$\sim 0.4-1.5$ TeV.
Therefore, with this very simple extension it is possible
to obtain a LWSM that can be tested at the LHC.
A careful study of this scenario and other possible extensions beyond
the minimal LWSM is needed.

\section*{Acknowledgments}
L. D. thanks Gustavo Burdman for many useful
discussions.
We thank Martin Gr\"{u}newald for providing us
the contours in the $S,T$ plane. 
L. D. acknowledges the financial support of the State of S\~ao
Paulo Research Foundation (FAPESP).

\appendix

\section{Diagonalization of the fermionic mass matrix}\label{appendixMf}
We consider the diagonalization of the fermionic mass matrix of the
third generation. We collect the up fermions into a three dimensional
vector in the following way:
\begin{equation}\label{fbasis}
\psi^{u\;t}_L=(q^u_L,\tilde q^u_L,\tilde u_L)\qquad 
\psi^{u\;t}_R=(u_R,\tilde u_R,\tilde q^u_R)\; ,
\end{equation}
and similarly for the down fermions. We adopt the same basis as
\cite{Krauss:2007bz}, but with a different notation. Using
Eq.~(\ref{fbasis}) we can write the quadratic fermionic Lagrangian
(\ref{eqFermion2}) as:
\begin{equation}\label{eqFermion2bis}
\mathcal{L}_{2\psi}=\bar\psi^u i\dslash\eta\psi^u-
\bar\psi^u_R \mathcal{M}_u \eta \psi^u_L-\bar\psi^u_L \eta \mathcal{M}^{\dagger}_u \psi^u_R+\dots \; ,
\end{equation}
where the dots stand for the down sector, $\eta=\rm{diag}(1,-1,-1)$ 
and 
\begin{equation}\label{Mfermion}
\mathcal{M}_u \eta=
\left(
 \begin{array}{ccc}
   m_u\;&\;-m_u\;&\;0 \\
   -m_u\;&\;m_u\;&\;-M_u \\
   0\;&\;-M_q\;&\;0 \\
    \end{array}\right)
\end{equation}

The mass matrix $\mathcal{M}_u$ can be diagonalized by independent left and
right symplectic rotations $S_{L,R}$ satisfying:
\begin{eqnarray} \label{ecs}
\mathcal{M}_{u,\mbox{{\tiny phys}}} \eta = S_R^\dagger \mathcal{M}_u \eta S_L \;,\qquad 
S_R \eta S_R^\dagger=\eta\;,\qquad 
S_L \eta S_L^\dagger = \eta\;,
\end{eqnarray}
where $\mathcal{M}_{u,\mbox{{\tiny phys}}}$ is the physical mass
matrix, which is diagonal. 

To obtain explicit analytic expressions we expand the solutions in powers of
Yukawa insertions $m_{u,d}$. Thus our results are well approximated by
the first terms in this expansion in the limit 
$\epsilon_{q,u}=\frac{m_u}{M_{q,u}}\ll 1$. For the elements of the
diagonal matrix $\mathcal{M}_{u,\mbox{{\tiny phys}}}$ we obtain the following:
\begin{eqnarray}\label{fmasses}
&&m_u[1+\frac{1}{2}(\epsilon_q^2+\epsilon_u^2)+\frac{1}{8}(7\epsilon_q^4+7\epsilon_u^4+10\epsilon_q^2\epsilon_u^2)]+
\mathcal{O}(\epsilon_{q,u}^6)\; ,\\
&&M_u[1-\frac{\epsilon_u^2}{2}\frac{M_q^2}{M_q^2-M_u^2}-
\frac{\epsilon_u^4}{8}\frac{5M_q^6-9M_q^4M_u^2}{(M_q^2-M_u^2)^3}]+
\mathcal{O}(\epsilon_{q,u}^6)\; ,\\
&&M_q[1+\frac{\epsilon_q^2}{2}\frac{M_u^2}{M_q^2-M_u^2}+
\frac{\epsilon_q^4}{8}\frac{5M_u^6-9M_u^4M_q^2}{(M_q^2-M_u^2)^3}]+
\mathcal{O}(\epsilon_{q,u}^6)\; .
\end{eqnarray}
For the matrices $S_{L,R}$ we obtain
\begin{eqnarray}\label{frotationl}
S_L-1=
\end{eqnarray}
\begin{eqnarray}
\left[
 \begin{array}{ccc}
\frac{\epsilon_u^2}{2} + 
   \frac{4\,\epsilon_q^4 + 8\,\epsilon_q^2\,\epsilon_u^2 + 
      11\,\epsilon_u^4}{8} & 
\frac{-\left( \epsilon_q^8\,
        \left( 4\,\epsilon_q^2 - 7\,\epsilon_u^2 \right)  \right) 
      }{2\,{\left( \epsilon_q^2 - \epsilon_u^2 \right) }^3} + 
   \frac{\epsilon_q^4}{-\epsilon_q^2 + \epsilon_u^2} & 
     -\epsilon_u - \frac{\epsilon_u^5\,
      \left( -4\,\epsilon_q^2 + 3\,\epsilon_u^2 \right) }{2\,
      {\left( \epsilon_q^2 - \epsilon_u^2 \right) }^2}             \\
-{\epsilon_q}^2 - \frac{\epsilon_q^2\,
      \left( 4\,\epsilon_q^2 + 3\,\epsilon_u^2 \right) }{2} &
   \frac{-\left( \epsilon_q^4\,\epsilon_u^2 \right) }
    {2\,{\left( \epsilon_q^2 - \epsilon_u^2 \right) }^2} + 
   \frac{\epsilon_q^8\,\left( 4\,\epsilon_q^4 - 
        16\,\epsilon_q^2\,\epsilon_u^2 + 23\,\epsilon_u^4
        \right) }{8\,{\left( \epsilon_q^2 - \epsilon_u^2 \right)
          }^4} & 
\frac{\epsilon_q^2\,\epsilon_u\,
      \left( -2\,\epsilon_q^4 + 
        4\,\epsilon_q^2\,\epsilon_u^2 - 2\,\epsilon_u^4
        \right) }{2\,{\left( \epsilon_q^2 - \epsilon_u^2 \right)
          }^3} + \frac{\epsilon_q^2\,\epsilon_u\,
      \left( 4\,\epsilon_q^2\,\epsilon_u^4 - 
        \epsilon_u^6 \right) }{2\,
      {\left( \epsilon_q^2 - \epsilon_u^2 \right) }^3}          \\
-\epsilon_u - \frac{\epsilon_u\,
      \left( 2\,\epsilon_q^2 + 3\,\epsilon_u^2 \right) }{2} &
   \frac{\epsilon_q^6\,\epsilon_u\,
      \left( 2\,\epsilon_q^2 - 5\,\epsilon_u^2 \right) }{2\,
      {\left( \epsilon_q^2 - \epsilon_u^2 \right) }^3} + 
   \frac{\epsilon_q^2\,\epsilon_u}
    {\epsilon_q^2 - \epsilon_u^2} & \frac{\epsilon_u^4\,
      \left( -2\,\epsilon_q^2 + \epsilon_u^2 \right) }{2\,
      {\left( \epsilon_q^2 - \epsilon_u^2 \right) }^2} + 
   \frac{\epsilon_u^8\,\left( 36\,\epsilon_q^4 - 
        36\,\epsilon_q^2\,\epsilon_u^2 + 11\,\epsilon_u^4
        \right) }{8\,{\left( \epsilon_q^2 - \epsilon_u^2 \right)
          }^4}
 \end{array}
\right]
\nonumber
\end{eqnarray}

\begin{eqnarray}\label{frotation}
S_R-1=
\end{eqnarray}
\begin{eqnarray}
\left[
 \begin{array}{ccc}
 \frac{\epsilon_q^2}{2} + 
   \frac{11\,\epsilon_q^4 + 8\,\epsilon_q^2\,\epsilon_u^2 + 
      4\,\epsilon_u^4}{8} & -\epsilon_q - 
   \frac{\epsilon_q^5\,\left( 3\,\epsilon_q^2 - 
        4\,\epsilon_u^2 \right) }{2\,
      {\left( \epsilon_q^2 - \epsilon_u^2 \right) }^2} & \frac
      {\epsilon_u^4}{\epsilon_q^2 - \epsilon_u^2} + 
   \frac{\epsilon_u^8\,\left( -7\,\epsilon_q^2 + 
        4\,\epsilon_u^2 \right) }{2\,
      {\left( \epsilon_q^2 - \epsilon_u^2 \right) }^3} \\
 -{
       \epsilon_u}^2 - \frac{\epsilon_u^2\,
      \left( 3\,\epsilon_q^2 + 4\,\epsilon_u^2 \right) }{2} & 
     \frac{\epsilon_q\,\epsilon_u^2\,
      \left( \epsilon_q^6 - 
        4\,\epsilon_q^4\,\epsilon_u^2 \right) }{2\,
      {\left( \epsilon_q^2 - \epsilon_u^2 \right) }^3} + 
   \frac{\epsilon_q\,\epsilon_u^2\,
      \left( 2\,\epsilon_q^4 - 
        4\,\epsilon_q^2\,\epsilon_u^2 + 2\,\epsilon_u^4
        \right) }{2\,{\left( \epsilon_q^2 - \epsilon_u^2 \right) }^
       3} & \frac{-\left( \epsilon_q^2\,\epsilon_u^4 \right) }
    {2\,{\left( \epsilon_q^2 - \epsilon_u^2 \right) }^2} + 
   \frac{\epsilon_u^8\,\left( 23\,\epsilon_q^4 - 
        16\,\epsilon_q^2\,\epsilon_u^2 + 4\,\epsilon_u^4
        \right) }{8\,{\left( \epsilon_q^2 - \epsilon_u^2 \right) }^
       4} \\
 -\epsilon_q - 
   \frac{\epsilon_q\,\left( 3\,\epsilon_q^2 + 
        2\,\epsilon_u^2 \right) }{2} & \frac{\epsilon_q^4\,
      \left( \epsilon_q^2 - 2\,\epsilon_u^2 \right) }{2\,
      {\left( \epsilon_q^2 - \epsilon_u^2 \right) }^2} + 
   \frac{\epsilon_q^8\,\left( 11\,\epsilon_q^4 - 
        36\,\epsilon_q^2\,\epsilon_u^2 + 36\,\epsilon_u^4
        \right) }{8\,{\left( \epsilon_q^2 - \epsilon_u^2 \right) }^
       4} & \frac{\epsilon_q\,\epsilon_u^6\,
      \left( 5\,\epsilon_q^2 - 2\,\epsilon_u^2 \right) }{2\,
      {\left( \epsilon_q^2 - \epsilon_u^2 \right) }^3} - 
   \frac{\epsilon_q\,\epsilon_u^2}
    {\epsilon_q^2 - \epsilon_u^2}
 \end{array}
\right]
\nonumber
\end{eqnarray}
The solution for the down-sector can be obtained from the up-sector
simply by changing $u\rightarrow d$.

The authors of Ref.~\cite{Krauss:2007bz} considered the case
$M_q=M_u$. Their solutions can not be obtained from the case $M_q\neq
M_u$ that is singular in the limit $M_q\rightarrow M_u$. There is a
singularity because in that limit there is a degenerate eigenspace of 
dimension two, with no preferred eigenvectors.

\section{Mass insertion resummation of the flavor propagators}\label{Approp}
In this Appendix we provide expressions for the flavor propagators to
all orders in the mass insertions. Due to the mixings between SM and
LW-fermions, the fermionic propagators are also mixed in the flavor
basis in the mass insertion expansion. The resummation of the mass
insertion series can be performed and the flavor propagators shown in
Eq.~(\ref{Stildeqs}) are those used in Sections~\ref{fermionicT}
and~\ref{fermionicS} to compute the full radiative fermionic
contributions to $T$ and $S$ respectively. We illustrate the method for
resumming the mass insertion series with a particular example. Other
cases are simple variations of the one discussed below and they can be
obtained by using the same procedure.

Consider the case of the resummed propagator ($\tilde S_{q^u}$) of a
SM up-fermion in a SU(2)$_L$ doublet. The first three terms of the
series are shown in Fig.~\ref{Sqprop}. The first term corresponds to the
zeroth order propagator ($S_{q}$) which is obtained from the SM
kinetic term in the Lagrangian given by Eq.~(\ref{eqFermion2}):
\begin{eqnarray} \label{kinetic}
\mathcal{L}_{2\psi}&\supset&\bar q_L i\dslash q_L \;,
\end{eqnarray}
and it takes the form:
\begin{eqnarray} \label{Sq}
\tilde S_{q^u}^{(0)} &\equiv& S_{q} = P_L \frac{1}{\pslash} \;,
\end{eqnarray}
where $P_L=(1-\gamma^5)/2$.
\begin{figure}
\begin{center}
\begin{picture}(410,60)(0,0)
        \LongArrow(0,35)(28,35)
        \Text(28,15)[c]{$\tilde S_{q^u}$}
        \Line(28,35)(50,35)
        \Text(57,35)[c]{=}
        \Line(70,35)(112,35)
        \Text(87,15)[c]{$S_q$}
        \Text(120,35)[c]{+}
        \Line(127,35)(160,35) 
        \Text(145,15)[c]{$S_q$}
        \Vertex(160,35){2}
        \Line(160,35)(190,35)
        \Text(175,15)[c]{$_{u- \tilde u}$}
        \Vertex(190,35){2}
        \Line(190,35)(220,35)
        \Text(205,15)[c]{$S_q$}
        \Text(230,35)[c]{+}
        \Line(235,35)(385,35)
        \Text(250,15)[c]{$S_q$}
        \Vertex(265,35){2}
        \Text(280,15)[c]{$_{u- \tilde u}$}
        \Vertex(295,35){2}
        \Text(310,15)[c]{$_{q- \qt}$}
        \Vertex(325,35){2}
        \Text(340,15)[c]{$_{u- \tilde u}$}
        \Vertex(355,35){2}
        \Text(370,15)[c]{$S_q$}
        \Text(395,35)[c]{+}
        \Text(408,35)[c]{$\dots$}        
\end{picture}
\caption{Expansion in mass insertions of the propagator ($\tilde
S_{q^u}$) of a SM up-fermion in the SU(2)$_L$ doublet. 
$S_q$ stands for the $q^u$ propagator with no mass
insertions. Besides, $u - \tilde u$ ($q - \qt$) corresponds to
internal zeroth order propagators for $u_R$ ($q^u$) and $\tilde u_R$
($\qt$).}
\label{Sqprop}
\end{center}
\end{figure}
 The following two terms in the expansion, containing at least two mass
 insertions, can be derived by taking into account the mixing mass
 term:
\begin{eqnarray} \label{mixing}
\mathcal{L}_{2\psi}&\supset&m_u(\bar{u}_R-\bar{\tilde{u}}_R)(q_L^u-\tilde
q_L^u) \;.
\end{eqnarray}
The first and second order propagators in mass insertions are found to
be respectively:
\begin{eqnarray} \label{23}
\hspace*{-0.7cm} \tilde S_{q^u}^{(1)} &=& m_u^2 P_L S_q (S+S_{\tilde
u}) P_L S_q = m_u^2 S_q \pslash P_L S_q A_u \;\;,\qquad \nonumber\\
\hspace*{-0.7cm} \tilde S_{q^u}^{(2)} &=& m_u^2 P_L S_q (S+S_{\tilde
u}) m_u (S_q+S_{\qt}) m_u (S+S_{\tilde u}) P_L S_q = m_u^2 S_q \pslash
P_L S_q A_u (m_u^2 p^2 A_q A_u)\;,
\end{eqnarray}
where $A_u$ and $A_q$, and the zeroth order propagators for $u_R$,
$\tilde{u}_R$ and $\qt$ ($S$, $S_{\tilde{u}}$ and $S_{\qt}$
respectively) are given by:
\begin{eqnarray} \label{As}
A_u &\equiv& \frac{1}{p^2}-\frac{1}{p^2-M^2_{u}}\;\;,\qquad A_q \equiv
\frac{1}{p^2}-\frac{1}{p^2-M^2_{q}}\;, \nonumber\\  S &\equiv& P_R
\frac{1}{\pslash} \;\;,\qquad\qquad S_{\tilde u} \equiv
-\frac{1}{\pslash+M_{u}}\;\;,\qquad\qquad S_{\qt} \equiv
-\frac{1}{\pslash+M_{q}}\;,
\end{eqnarray}
with $P_R=(1+\gamma^5)/2$. Notice the extra
minus sign and the absence of any chirality projector in $S_{\qt}$ and
$S_{\tilde u}$.
From the results in Eq.~(\ref{23}), it is not difficult to infer the
generic $n$-th term and the sum of the series can be obtained:
\begin{eqnarray} \label{resum}
\tilde S_{q^u} &=& S_q +  m_u^2 S_q \pslash P_L S_q A_u
\sum_{n=0}^{\infty} {(m_u^2 p^2 A_q A_u)}^n = S_q + \frac{ m_u^2 S_q
\pslash P_L S_q A_u}{1 - m_u^2 p^2 A_q A_u}
\end{eqnarray}
Note that $\tilde S_{q^u}$ has only an even number of mass insertions.

Following similar arguments, it is possible to obtain all the resummed
flavor propagators. For SM and LW fermions of the third generation,
charged under SU(2)$_L$, there are four different types of propagators
arising from an even number of mass insertions. Initial and final legs
carry the same SM ($\tilde{S}_{q}$) or LW-fermion ($\tilde{S}_{\qt}$),
or they are attached to different fermions: the initial leg is a SM
($\tilde{S}_{q \qt}$) or a LW-fermion ($\tilde{S}_{\qt q}$) --the
subscript $q$ ($\qt$) stands for up or down SM (LW) fermions in the
SU(2)$_L$ doublet. This class of propagators enters the calculation
of both $S$ and $T$ parameters. The computation of the vacuum polarization
contributions to $S$ also requires propagators with an odd number of
mass insertions that can be obtained with the same method outlined
above. There are four relevant types of them according to all possible
combinations of SM and LW fermions in the SU(2)$_L$ doublet and up or
down singlets coupled to hypercharge ($\tilde M_{q (u,d)}, \tilde M_{q
(\tilde u, \tilde d)}, \tilde M_{\qt (\tilde u, \tilde d)}$ and
$\tilde M_{\qt (u,d)}$). Those obtained by interchanging initial and
final legs ($\tilde M_{ij} \rightarrow \tilde M_{ji}$) are needed as
well. Resumming all possible insertions of mixing mass terms, we
obtain the following expressions for propagators in the
up-sector:
\begin{eqnarray} \label{Stildeqs}
& &\tilde{S}_{q^u}(p) = S_{q} + \frac{m_u^2 S_{q} \pslash P_L S_{q}
A_u}{1-m_u^2 p^2 A_{q} A_u}\;\;,\qquad \tilde{S}_{\qt^u}(p) = S_{\qt}
+ \frac{m_u^2 S_{\qt} \pslash P_L S_{\qt} A_u}{1-m_u^2 p^2 A_{q}
A_u}\;, \nonumber\\  & &\tilde{S}_{q^u \qt^u}(p) = -\frac{m_u^2 S_{q}
\pslash P_L S_{\qt} A_u}{1-m_u^2 p^2 A_{q} A_u}\;\;,\qquad
\tilde{S}_{\qt^u q^u}(p) = -\frac{m_u^2 S_{\qt} \pslash P_L S_{q}
A_u}{1-m_u^2 p^2 A_{q} A_u}\;, \nonumber\\ & &\tilde{M}_{q^u u}(p) =-
\frac{m_u S_{q} P_R S}{1-m_u^2 p^2 A_{q} A_u}\;\;, \qquad\quad
\tilde{M}_{q^u \tilde u}(p) = \frac{m_u S_{q} P_R S_{\tilde
u}}{1-m_u^2 p^2 A_{q} A_u}\;, \nonumber\\ & &\tilde{M}_{\qt^u u}(p) =
\frac{m_u S_{\tilde q} P_R S}{1-m_u^2 p^2 A_{q} A_u}\;\;, \qquad
\tilde{M}_{\qt^u \tilde u}(p) =-\frac{m_u S_{\tilde q} P_R S_{\tilde
u}}{1-m_u^2 p^2 A_{q} A_u}\;,
\end{eqnarray}
where $S_q$, $S$, $S_{\tilde{u}}$,
$S_{\qt}$, $A_u$ and $A_{q}$ have been defined in
Eqs.~(\ref{Sq}) and (\ref{As}). $\tilde{M}_{ji}$ has a similar expression to
$\tilde{M}_{ij}$, only the place of the zeroth order propagators must
be interchanged as in the case of $\tilde{S}_{q^u \qt^u}$ and
$\tilde{S}_{\qt^u q^u}$. Note that the series for $\tilde{M}_{ij}$ and
$\tilde{S}_{ij}$ start from one and two mass insertions respectively
since the kinetic terms are flavor diagonal.

The fermionic spectrum of the up-sector is given by the poles of $\tilde{S}_{q^u}$.

Flavor propagators for the down-sector are obtained from those above
by changing $u \rightarrow d$ in Eqs.~(\ref{Stildeqs}).

\section{Fermionic contribution to the vacuum polarization}\label{Appvacuum}

We show in this Appendix the fermionic contributions to the vacuum
polarization associated to the $S$ and $T$ parameters:
\begin{equation}
S=\frac{16\pi}{g_1 g_2}\Pi_{3B}'(0)\; , \qquad
T=\frac{4\pi}{g_2^2 s^2 c^2 m_Z^2}[\Pi_{11}(0)-\Pi_{33}(0)]\; ,
\end{equation}
with $\Pi_{\mu\nu}=g_{\mu\nu}\Pi+(q_\mu q_\nu {\rm terms})$.

We consider first the perturbative expansion of $\Pi_{33}$ from 
Fig.~\ref{figLWfermions}(c). Since $u_R,d_R$ and their LW-partners are
singlets of SU(2)$_L$, and the gauge interactions do not mix SM and
LW-fermions, the fermionic legs attached to one of the gauge vertices
correspond either to $q$ or to $\qt$. However, it is possible to have
$q$-legs attached to one of the vertices and either $q$ or $\qt$-legs
attached to the other vertex, and similar for $\qt$. Using the
propagators of Appendix~\ref{Approp} we can write the up contribution
to $\Pi_{33}$ to all orders in the mass insertion expansion as:
\begin{equation}
\Pi_{33}^{\mu\nu}=-\frac{g_2^2}{4}{\rm tr}\int\frac{d^4p}{(2\pi)^4}
(\gamma^\mu \tilde S_{q^u}\gamma^\nu \tilde S_{q^u}+\gamma^\mu \tilde
S_{\qt^u}\gamma^\nu \tilde S_{\qt^u}-2\gamma^\mu \tilde
S_{q^u\qt^u}\gamma^\nu \tilde S_{\qt^u q^u}),
\end{equation}
and a similar contribution from the down sector. The minus sign in the
last term is because the LW-fermions couple to the SM-gauge fields
with a sign flip compared with the SM-fermions, see Eq.~(\ref{eqLint}).

The contributions to $\Pi_{11}$ can be obtained in a similar way, and
is given by:
\begin{equation}
\Pi_{11}^{\mu\nu}=-\frac{g_2^2}{2}{\rm tr}\int\frac{d^4p}{(2\pi)^4}
(\gamma^\mu \tilde S_{q^u}\gamma^\nu \tilde S_{q^d}+\gamma^\mu \tilde
S_{\qt^u}\gamma^\nu \tilde S_{\qt^d}-2\gamma^\mu \tilde
S_{q^u\qt^u}\gamma^\nu \tilde S_{\qt^d q^d}).
\end{equation}

The $S$ parameter is proportional to $\Pi'_{3B}(0)$. The fermionic
contribution to $\Pi_{3B}$ is more involved because the fermions
$u_R,d_R$ and their LW-partners couple to hypercharge. Therefore we
have to consider the diagrams of Figs.~\ref{figLWfermions}(b) and
(c). The contribution from Fig.~\ref{figLWfermions}(c) is similar to
$\Pi_{33}$, with the appropriate charges:
\begin{equation}
\Pi_{3B}^{\mu\nu}=-\frac{g_1g_2}{12}{\rm tr}\int\frac{d^4p}{(2\pi)^4}
(\gamma^\mu \tilde S_{q^u}\gamma^\nu \tilde S_{q^u}+\gamma^\mu \tilde
S_{\qt^u}\gamma^\nu \tilde S_{\qt^u}-2\gamma^\mu \tilde
S_{q^u\qt^u}\gamma^\nu \tilde S_{\qt^u q^u}),
\end{equation}
and a similar contribution from the down sector with a minus sign due
to the different weak charge.

Fig.~\ref{figLWfermions}(b) gives contributions with $q,\qt$-legs
attached to $W_3$ and the singlets $u_R,d_R,\ut,\dt$ attached to $B$. 
Using the propagators of Appendix~\ref{Approp}, the up contribution
to $\Pi_{3B}$, to all orders in the mass insertion expansion, is:
\begin{equation}\label{Suqu}
\Pi_{3B}^{\mu\nu}=-\frac{g_1g_2}{3}{\rm tr}\int\frac{d^4p}{(2\pi)^4}
(\gamma^\mu \tilde M_{q^uu}\gamma^\nu \tilde M_{uq^u}-
\gamma^\mu \tilde M_{q^u\ut}\gamma^\nu \tilde M_{\ut q^u}-
\gamma^\mu \tilde M_{\qt^u u}\gamma^\nu \tilde M_{u\qt^u}+
\gamma^\mu \tilde M_{\qt^u\ut}\gamma^\nu \tilde M_{\ut\qt^u}).
\end{equation}
The contribution from the down sector can be obtained from
Eq.~(\ref{Suqu}) by changing the factor $\frac{1}{3}$ by
$\frac{1}{6}$ and changing the indices $u\rightarrow d$.

%\newpage

{}


\begin{thebibliography}{}

%\cite{Grinstein:2007mp}
\bibitem{Grinstein:2007mp}
  B.~Grinstein, D.~O'Connell and M.~B.~Wise,
  %``The Lee-Wick standard model,''
  Phys.\ Rev.\  D {\bf 77} (2008) 025012
  [arXiv:0704.1845 [hep-ph]].
  %%CITATION = PHRVA,D77,025012;%%

%\cite{Lee:1969fy}
\bibitem{Lee:1969fy}
  T.~D.~Lee and G.~C.~Wick,
  %``Negative Metric and the Unitarity of the S Matrix,''
  Nucl.\ Phys.\  B {\bf 9}, 209 (1969).
  %%CITATION = NUPHA,B9,209;%%

%\cite{Lee:1970iw}
\bibitem{Lee:1970iw}
  T.~D.~Lee and G.~C.~Wick,
  %``Finite Theory of Quantum Electrodynamics,''
  Phys.\ Rev.\  D {\bf 2}, 1033 (1970).
  %%CITATION = PHRVA,D2,1033;%%

%\cite{Grinstein:2007iz}
\bibitem{Grinstein:2007iz}
  B.~Grinstein, D.~O'Connell and M.~B.~Wise,
  %``Massive Vector Scattering in Lee-Wick Gauge Theory,''
  arXiv:0710.5528 [hep-ph].
  %%CITATION = ARXIV:0710.5528;%%

%\cite{Cutkosky:1969fq}
\bibitem{Cutkosky:1969fq}
  R.~E.~Cutkosky, P.~V.~Landshoff, D.~I.~Olive and J.~C.~Polkinghorne,
  %``A non-analytic S matrix,''
  Nucl.\ Phys.\  B {\bf 12}, 281 (1969).
  %%CITATION = NUPHA,B12,281;%%

%\cite{Nakanishi:1971jj}
\bibitem{Nakanishi:1971jj}
  N.~Nakanishi,
  %``Lorentz noninvariance of the complex-ghost relativistic field theory,''
  Phys.\ Rev.\  D {\bf 3}, 811 (1971).
  %%CITATION = PHRVA,D3,811;%%

%\cite{Lee:1971ix}
\bibitem{Lee:1971ix}
  T.~D.~Lee and G.~C.~Wick,
  %``QUESTIONS OF LORENTZ INVARIANCE IN FIELD THEORIES WITH INDEFINITE METRIC,''
  Phys.\ Rev.\  D {\bf 3}, 1046 (1971).
  %%CITATION = PHRVA,D3,1046;%%

%\cite{Nakanishi:1971ky}
\bibitem{Nakanishi:1971ky}
  N.~Nakanishi,
  %``Remarks on the dipole-ghost scattering states,''
  Phys.\ Rev.\  D {\bf 3}, 1343 (1971).
  %%CITATION = PHRVA,D3,1343;%%

%\cite{Antoniadis:1986tu}
\bibitem{Antoniadis:1986tu}
  I.~Antoniadis and E.~T.~Tomboulis,
  %``Gauge Invariance And Unitarity In Higher Derivative Quantum Gravity,''
  Phys.\ Rev.\  D {\bf 33}, 2756 (1986).
  %%CITATION = PHRVA,D33,2756;%%

%\cite{Boulware:1983vw}
\bibitem{Boulware:1983vw}
  D.~G.~Boulware and D.~J.~Gross,
  %``Lee-Wick Indefinite Metric Quantization: A Functional Integral Approach,''
  Nucl.\ Phys.\  B {\bf 233}, 1 (1984).
  %%CITATION = NUPHA,B233,1;%%

%\cite{Jansen:1993jj}
\bibitem{Jansen:1993jj}
  K.~Jansen, J.~Kuti and C.~Liu,
  %``The Higgs model with a complex ghost pair,''
  Phys.\ Lett.\  B {\bf 309}, 119 (1993)
  [arXiv:hep-lat/9305003].
  %%CITATION = PHLTA,B309,119;%%

%\cite{Jansen:1993ji}
\bibitem{Jansen:1993ji}
  K.~Jansen, J.~Kuti and C.~Liu,
  %``Strongly interacting Higgs sector in the minimal Standard Model?,''
  Phys.\ Lett.\  B {\bf 309}, 127 (1993)
  [arXiv:hep-lat/9305004].
  %%CITATION = PHLTA,B309,127;%%

%\cite{Fodor:2007fn}
\bibitem{Fodor:2007fn}
  Z.~Fodor, K.~Holland, J.~Kuti, D.~Nogradi and C.~Schroeder,
  %``New Higgs physics from the lattice,''
  arXiv:0710.3151 [hep-lat].
  %%CITATION = NONE,,;%%

%\cite{Knechtli:2007ea}
\bibitem{Knechtli:2007ea}
  F.~Knechtli, N.~Irges and M.~Luz,
  %``New Higgs mechanism from the lattice,''
  arXiv:0711.2931 [hep-ph].
  %%CITATION = ARXIV:0711.2931;%%

%\cite{Grinstein:2008qq}
\bibitem{Grinstein:2008qq}
  B.~Grinstein and D.~O'Connell,
  %``One-Loop Renormalization of Lee-Wick Gauge Theory,''
  arXiv:0801.4034 [hep-ph].
  %%CITATION = ARXIV:0801.4034;%%

%\cite{Dulaney:2007dx}
\bibitem{Dulaney:2007dx}
  T.~R.~Dulaney and M.~B.~Wise,
  %``Flavor Changing Neutral Currents in the Lee-Wick Standard Model,''
  arXiv:0708.0567 [hep-ph].
  %%CITATION = ARXIV:0708.0567;%%

%\cite{Wu:2007yd}
\bibitem{Wu:2007yd}
  F.~Wu and M.~Zhong,
  %``The Lee-Wick Fields out of Gravity,''
  Phys.\ Lett.\  B {\bf 659} (2008) 694
  [arXiv:0705.3287 [hep-ph]].
  %%CITATION = PHLTA,B659,694;%%

%\cite{Espinosa:2007ny}
\bibitem{Espinosa:2007ny}
  J.~R.~Espinosa, B.~Grinstein, D.~O'Connell and M.~B.~Wise,
  %``Neutrino masses in the Lee-Wick standard model,''
  arXiv:0705.1188 [hep-ph].
  %%CITATION = ARXIV:0705.1188;%%

%\cite{Rizzo:2007ae}
\bibitem{Rizzo:2007ae}
  T.~G.~Rizzo,
  %``Searching for Lee-Wick Gauge Bosons at the LHC,''
  JHEP {\bf 0706}, 070 (2007)
  [arXiv:0704.3458 [hep-ph]].
  %%CITATION = JHEPA,0706,070;%%

%\cite{Krauss:2007bz}
\bibitem{Krauss:2007bz}
  F.~Krauss, T.~E.~J.~Underwood and R.~Zwicky,
  %``The process gg -> h_0 -> gamma gamma in the Lee-Wick Standard Model,''
  arXiv:0709.4054 [hep-ph].
  %%CITATION = ARXIV:0709.4054;%%

%\cite{Rizzo:2007nf}
\bibitem{Rizzo:2007nf}
  T.~G.~Rizzo,
  %``Unique Identification of Lee-Wick Gauge Bosons at Linear Colliders,''
  JHEP {\bf 0801} (2008) 042
  [arXiv:0712.1791 [hep-ph]].
  %%CITATION = JHEPA,0801,042;%%

\bibitem{lepparadox}
  R.~Barbieri and A.~Strumia,
  %``The 'LEP paradox',''
  arXiv:hep-ph/0007265.
  %%CITATION = HEP-PH/0007265;%%

\bibitem{oblique}
  M.~E.~Peskin and T.~Takeuchi,
  %``A New constraint on a strongly interacting Higgs sector,''
  Phys.\ Rev.\ Lett.\  {\bf 65} (1990) 964;
  %%CITATION = PRLTA,65,964;%%
  %M.~E.~Peskin and T.~Takeuchi,
  %``Estimation of oblique electroweak corrections,''
  Phys.\ Rev.\  D {\bf 46} (1992) 381.
  %%CITATION = PHRVA,D46,381;%%

\bibitem{MFV}
  R.~S.~Chivukula and H.~Georgi,
  %``Composite Technicolor Standard Model,''
  Phys.\ Lett.\  B {\bf 188} (1987) 99;
  %%CITATION = PHLTA,B188,99;%%
  G.~D'Ambrosio, G.~F.~Giudice, G.~Isidori and A.~Strumia,
  %``Minimal flavour violation: An effective field theory approach,''
  Nucl.\ Phys.\  B {\bf 645} (2002) 155
  [arXiv:hep-ph/0207036].
  %%CITATION = NUPHA,B645,155;%%

\bibitem{Agashe:2003zs}
  K.~Agashe, A.~Delgado, M.~J.~May and R.~Sundrum,
  %``RS1, custodial isospin and precision tests,''
  JHEP {\bf 0308} (2003) 050
  [arXiv:hep-ph/0308036].
  %%CITATION = JHEPA,0308,050;%%

\bibitem{EWWG}
J.~Alcaraz {\it et al.}  [LEP Collaboration],
  %``Precision Electroweak Measurements and Constraints on the Standard Model,''
  arXiv:0712.0929 [hep-ex].
  %%CITATION = ARXIV:0712.0929;%%


\end{thebibliography}
\end{document}